\documentclass{aastex}
\usepackage{spr-astr-addons}
\usepackage{url}

\RequirePackage{color}

\begin{document}

\title{Singularity-free anisotropic strange quintessence star }
\shorttitle{Singularity-free anisotropic strange quintessence star }
\shortauthors{Piyali Bhar.}

\author{Piyali Bhar\altaffilmark{1,*}}

\altaffiltext{1}{ Department of
Mathematics, Jadavpur University, Kolkata 700 032, West Bengal,
India}
\altaffiltext{*}{Email:piyalibhar90@gmail.com}

\begin{abstract}
Present paper provides a new model of anisotropic strange star corresponding to the exterior schwarzschild metric.The Einstein field equations have been solved by utilizing the Krori-Barua (KB) ansatz [K.D. Krori and J. Barua, {\it J. Phys. A: Math. Gen.} {\bf 8}, 508 (1975)] in presence of quintessence field characterized by a parameter $\omega_q$ with $-1<\omega_q<-\frac{1}{3}$.The obtained solutions are free from central singularity. Our model is potentially stable.The numerical values of mass of the different strange stars SAXJ1808.4-3658(SS1)(radius=7.07 km),4U1820-30 (radius=10 km),Vela X-12 (radius=9.99 km),PSR~J 1614-2230 (radius=10.3 km) obtained from our model is very close to the observational data that confirms the validity of our proposed model. The interior solution is also matched to the exterior Schwarzschild spacetime in presence of thin shell where negative surface pressure is required to hold the thin shell against collapse.
\end{abstract}


\maketitle

\section{Introduction}
One of the most important discoveries in the last decades is that the expansion of our present universe is accelerating. It was first observed by high red shift supernova Ia and later it was confirmed by cosmic microwave radiation \citep{bennett03,spergel03}. Dark energy is the most suitable candidate to explain this. As a result the study of dark energy has become a subject of considerable interest to the researchers.Cosmic Microwave Background (CMB) has shown that our universe made up of 68.3\% dark energy,26.8\% dark matter and 4.9\% ordinary matter. Dark matter is attractive in nature and it can not be seen by telescope but its existence has been proved by gravitational effects on visible matter and gravitational lensing of background radiation. On the other hand dark energy is repulsive in nature and it has strong negative pressure. $p=\omega \rho$ with $\omega<0$ is generally called the dark energy equation of state. $\omega$ is called the dark energy parameter.For accelerating expansion $\omega $ should lie in the range $\omega<-\frac{1}{3}$. If $\omega$ lies in the range $-1<\omega<-\frac{1}{3}$ it is referred to as quintessence field. $\omega<-1$ is named as phantom regime that has a peculiar property namely infinitely increasing energy density. In particular if $\omega=-1$ then the dark energy equation of states becomes $p=-\rho$ which describes the equation of state of the shell of a 'Gravastar', gravitationally vacuum condense star proposed by Mazur and Mottola \citep{mazur01,mazur04}. Some other works on gravastar can be found in the references \citep{usmani11,far12a,far12b,piyali14c}. \\

To study stellar structure and evolution it is generally assumed that the underlying fluid is a perfect fluid,i.e,the pressure inside the fluid sphere is isotropic in nature. However present observation shows that the fluid pressure of the highly compact astrophysical objects like X-ray pulsar,Her-X-1, X-ray buster 4U 1820-30,millisecond pulsar SAXJ1804.4-3658 etc. whose density of core is expected to be beyond the nuclear density ($\sim 10^{15}$ gm/cc) becomes anisotropy in nature,i.e,it can be decomposed into two parts radial pressure $p_r$ and transverse pressure $p_t$ where $p_t$ is in the perpendicular direction to $p_r.$ $\Delta=p_t-p_r$ is called the anisotropic factor. Local anisotropy in self-gravitating systems were studied by Herrera and Santos \citep{herrera97} and see the the references there in for a review of anisotropic fluid sphere. Anisotropy may occurs in various reasons e.g,the existence of solid core,in presence of type P superfluid, phase transition,rotation, magnetic field,mixture of two fluid,existence of external field etc.It is believed that strange quark matter is consisted of $u,d$ and $s$ quarks. According to Witten \citep{witten84} the formation of strange matter can be classified into two ways:the quark hadron phase transition in the early universe and conversion of neutron stars into strange stars at ultrahigh densities. Stars composed of strange matter is called the strange star which can be classified into two types:for type-I strange star $\frac{m}{a}>0.3$ and for type-II strange star $0.2<\frac{m}{a}\leq 0.3$ and to distinguish the type-II strange stars from the neutron star the information about density profile,mass,radius are essential \citep{jotania06}. In the present paper we propose a model of strange star in presence of quintessence field. Since the density inside a strange star is beyond the nuclear density,we have considered the pressure anisotropy for our model.  We assume that the underlying fluid is a mixture of ordinary matter and an unknown matter which is of dark energy type i.e repulsive in nature. Dark energy star models have been studied by several authors.Lobo \citep{lobo06} has given a model of stable dark energy star by assuming two special type of mass function one is of constant energy density and the other mass function is Tolman-Whitker mass. All the features of the dark energy star has been discussed and the system is stable under small linear perturbation.The van der Waals quintessence stars have been studied by Lobo\citep{lobo07}.In that work, the construction of inhomogeneous compact spheres supported by a van der Waals equation of state is explored.van der Waals gravastar,van der Waals wormhole have also been discussed.Variable Equation of State for Generalized Dark Energy Model has been studied in \citep{saibal11}. Bhar and Rahaman \citep{piyali14} have proposed a new model of dark energy star consisting of three zones namely an inhomogeneous interior region with anisotropic pressures,thin shell and the exterior vacuum region of Schwarzschild spacetime.The proposed model satisfies all the physical requirements.The stability condition under small linear perturbation has also been discussed.Anisotropic Quintessence star has been studied by Kalam \emph{et al.}\citep{mehedi13}.Finch-skea ansatz \citep{finch89} was used to solve the Einstein field equation.The authors took a particular choice of the quintessence field to develop the model.In a very recent work Bhar \citep{bhar14b} has described one parameter group of conformal motion in presence of quintessence field.Vaidya-Titekar \citep{vaidya82} ansatz was used to develop the model.The obtained results are analyzed physically as well as with the help of graphical representation.\\

The paper is organized as follows: In sect.2 we have discussed about interior spacetime and Einstein field equations.Solutions of the system and physical analysis is done in sect.3 and sect.4 respectively.The other features are given in sect.5-9 and finally some concluding remarks is given in sect.10.

\section{Interior Solutions and Einstein field Equation}
To describe a static spherically symmetry spacetime let us consider the line element in the standard form as,
\begin{equation}
ds^{2}=-e^{\nu(r)}dt^{2}+e^{\lambda(r)}dr^{2}+r^{2}(d\theta^{2}+\sin^{2}\theta d\phi^{2})
\end{equation}
Where $\lambda$ and $\nu$ are functions of the radial parameter `r' only.\\
~~~~~~~Now let us assume that our model contains a quintessence like field along with anisotropic pressure representing normal matter.\\
The Einstein Equations can be written as,
\begin{equation}
G_{\mu \nu}=8\pi G (T_{\mu \nu}+\tau_{\mu \nu})
\end{equation}
Where $G_{\mu \nu}$ is the Einstein tensor and $T_{\mu\nu},\tau_{\mu \nu}$ are respectively the energy momentum tensor of the ordinary matter and quintessence like field characterized by a parameter $\omega_q$ with $-1<\omega_q<-\frac{1}{3}$. Now Kiselev \citep {kiselev03} has shown that the component of this tensor need to satisfy the conditions of additivity and linearity. Considering the different signature used in line elements,the components can be stated as follows:
\begin{equation}
\tau_t^{t}=\tau_r^{r}=-\rho_q
\end{equation}
\begin{equation}
\tau_{\theta}^{\theta}=\tau_{\phi}^{\phi}=\frac{1}{2}(3\omega_q+1)\rho_q
\end{equation}
and the corresponding energy-momentum tensor can be written as,
\begin{equation}
T_{\nu}^{\mu}=(\rho+p_r)u^{\mu}u_{\nu}-p_t g_{\nu}^{\mu}+(p_r-p_t)\eta^{\mu}\eta_{\nu}
\end{equation}
with $ u^{i}u_{j} =-\eta^{i}\eta_j = 1 $ and $u^{i}\eta_j= 0$. Here the vector $u_i$ is the fluid 4-velocity and $\eta^{i}$ is the spacelike vector which is orthogonal to $ u^{i}$, $\rho$ is the matter density, $p_r$ and $p_t$ are respectively the radial and the transversal pressure of the fluid.\\
The Einstein field equation assuming $G=1=c$ can be written as
\begin{equation}
e^{-\lambda}\left[\frac{\lambda'}{r}-\frac{1}{r^{2}} \right]+\frac{1}{r^{2}}=8\pi(\rho+\rho_q)
\end{equation}
\begin{equation}
e^{-\lambda}\left[\frac{1}{r^{2}}+\frac{\nu'}{r} \right]-\frac{1}{r^{2}}=8\pi(p_r-\rho_q)
\end{equation}

\[\frac{1}{2}e^{-\lambda}\left[ \frac{1}{2}\nu'^{2}+\nu''-\frac{1}{2}\lambda'\nu'+\frac{1}{r}(\nu'-\lambda')\right]\]
\begin{equation}
~~~~~~~~=8\pi\left(p_t+\frac{3\omega_q+1}{2}\rho_q
\right)
\end{equation}
\section{solution}
To solve the Einstein field equations we consider KB ansatz \citep{kb75}
\begin{equation}
\lambda(r)=Ar^{2},~~~~~~~~\nu(r)=Br^{2}+C
\end{equation}
Where $A,B$ and $C$ are some arbitrary constants which will be determined later using some physical conditions.\\
~~Since the pressure inside the fluid sphere is anisotropic in nature,so in this case $p_r \neq p_t$.\\
Now one can note that we have three equations (6)-(8) with four unknowns namely $\rho,p_r,p_t,\rho_q$.
To solve the above system of equations let us assume that the radial pressure $p_r$ is proportional to the matter density $\rho$ i.e,
\begin{equation}
p_r=\alpha\rho,~~~~~~~~~~~0<\alpha<1
\end{equation}
Where $\alpha$ is the equation of state parameter. The equation (10) corresponds to a polytropic equation of state of the second class. Where the polytropic constant is $\alpha$ and the polytropic index (n) is infinity ,a particular case of equation (36) in reference \citep{herrera13}\\
Solving equations $(6)-(8)$ with help of equations (9) and (10) one can obtain
\begin{equation}
\rho=\frac{A+B}{4\pi(1+\alpha)}e^{-Ar^{2}}
\end{equation}
\begin{equation}
p_r=\frac{\alpha(A+B)}{4\pi(1+\alpha)}e^{-Ar^{2}}
\end{equation}
\begin{equation}
\rho_q=\frac{e^{-Ar^{2}}}{8\pi}\left[2A-\frac{1}{r^{2}}-\frac{2(A+B)}{1+\alpha} \right]+\frac{1}{8\pi r^{2}}
\end{equation}
\begin{figure}
    \centering
        \includegraphics[scale=.3]{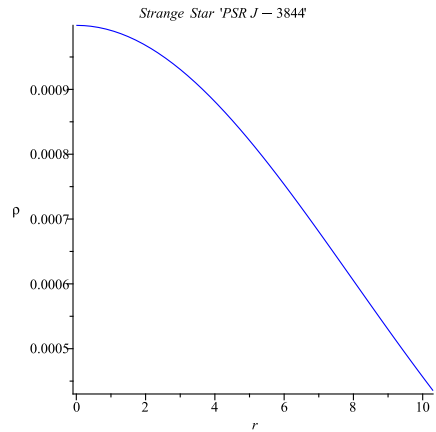}
    \caption{The variation of the matter density  $\rho$  vs r (in Km). }
    \label{Fig 5}
\end{figure}
\[p_t=\frac{e^{-Ar^{2}}}{8\pi}\left[B(B-A)r^{2}+(2B-A)\right]-\frac{3\omega_q+1}{16\pi}\times\]
\begin{equation}
~~~~~\left[e^{-Ar^{2}}\left\{2A-\frac{1}{r^{2}}-\frac{2(A+B)}{1+\alpha} \right\}+\frac{1}{r^{2}}\right]
\end{equation}
The profile of matter density $(\rho)$,radial pressure $(p_r)$ \& transverse pressure $p_t$ and quintessence field $(\rho_q)$  of strange star PSR~J 1614-2230 (Radius=10.3 km) are shown in fig.~1,fig~2 and fig.~3 respectively.

\begin{figure}
    \centering
       \includegraphics[scale=.3]{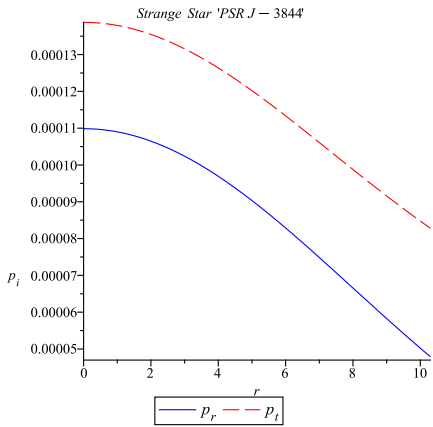}
    \caption{Radial pressure $p_r$ and Transverse pressure $p_t$ are plotted against `r' (in km) by taking $\alpha=0.11$ and $\omega_q=-0.7$ }
    \label{Fig 5}
\end{figure}

\section{Physical Analysis}
For a physical meaningful solution one must have
\begin{itemize}
  \item $\rho$, $ p_r $ and $p_t \geq 0$ for $0\leq r \leq a$
  \item matter density ($\rho$)and radial pressure ($p_r$) should be monotonic decreasing function of r
\end{itemize}
From the profile of $\rho,p_r,p_t$ given in Fig.~1 and Fig.~2 the first condition is satisfied.\\
The central density $\rho_0$ is given by,
\[\rho_0=\rho(r=0)=\frac{A+B}{4\pi(1+\alpha)}\]
and the surface density i.e, density on the boundary $r=a$ is given by,
\[\rho_a=\frac{A+B}{4\pi(1+\alpha)}e^{-Aa^{2}}\]
For our model,
\[\frac{d\rho}{dr}=-\frac{Are^{-Ar^{2}}}{2\pi (1+\alpha)}<0,~~~\frac{dp_r}{dr}=-\frac{\alpha Are^{-Ar^{2}}}{2\pi (1+\alpha)}<0\]
and at the point $r=0$,\\
\[\frac{d\rho}{dr}=0,~~\frac{dp_r}{dr}=0\]
also\\
 \[\frac{d^{2}\rho}{dr^{2}}=-\frac{A}{2\pi(1+\alpha)}<0,~~~\frac{d^{2}p_r}{dr^{2}}=-\frac{\alpha A}{2\pi(1+\alpha)}<0\]
\begin{figure}
    \centering
        \includegraphics[scale=.3]{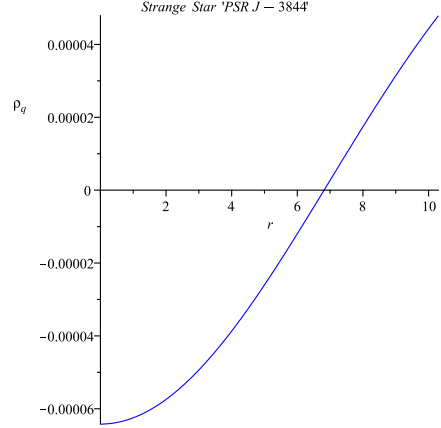}
    \caption{The plot of the quintessence field $\rho_q$ is plotted against `r' (km) by taking $\alpha=0.11$ and $\omega_q=-0.7$ }
    \label{Fig.5}
\end{figure}

So one can conclude that both $\rho$ and $p_r$ are monotonic decreasing function of r and they have maximum value at the center of the star.\\
The anisotropic force is given by $\Delta=\frac{2}{r}(p_t-p_r)$ which is depicted in fig.~4.From the figure we see that $\Delta>0$,i.e $p_t>p_r$ which concludes that the anisotropic force is repulsive in nature.

\begin{figure}
    \centering
        \includegraphics[scale=.3]{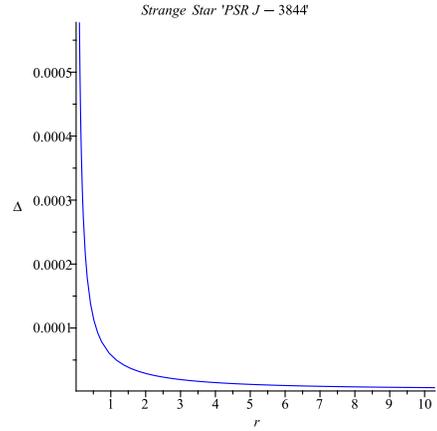}
    \caption{The anisotropic force is plotted against `r' by taking $\alpha=0.11$ and $\omega_q=-0.7$  }
    \label{Fig 5}
\end{figure}
The equation of state parameter $\omega_r$ and $\omega_t$ are obtained as,
\[\omega_r=\frac{p_r}{\rho}=\alpha\]
and
\[\omega_t=\frac{p_t}{\rho}=\]
\[\frac{1+\alpha}{2(A+B)}\left[B(B-A)r^{2}+(2B-A)\right]-\frac{3\omega_q+1}{4}\times\]
\[\frac{1+\alpha}{A+B}e^{Ar^{2}}\left[e^{-Ar^{2}}\left\{2A-\frac{1}{r^{2}}-\frac{2(A+B)}{1+\alpha} \right\}+\frac{1}{r^{2}}\right]\]
\begin{figure}
    \centering
        \includegraphics[scale=.3]{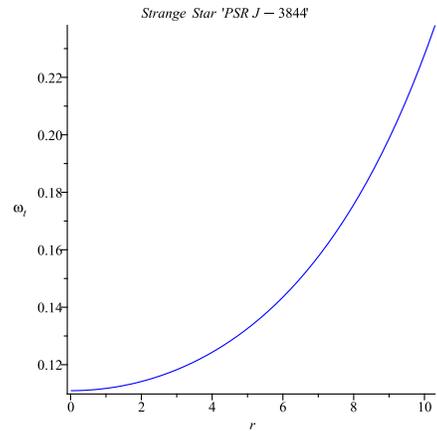}
    \caption{The equation of state parameter $\omega_t$ is plotted against `r' (km) by taking $\alpha=0.11$ and $\omega_q=-0.7$ }
    \label{Fig.5}
\end{figure}

\begin{figure}
    \centering
        \includegraphics[scale=.3]{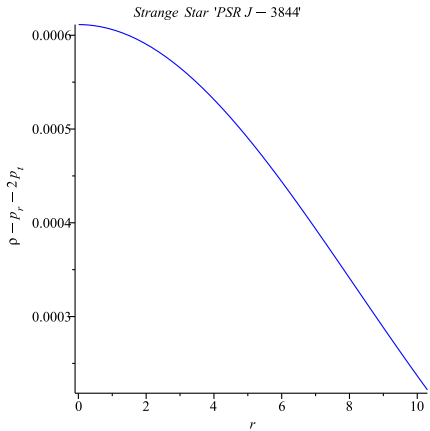}
    \caption{Trace of the energy momentum tensor is plotted against `r' (km) by taking $\alpha=0.11$ and $\omega_q=-0.7$ }
    \label{Fig.5}
\end{figure}
The profile of $\omega_t$ is shown in fig~5. From the figure it is clear that $\omega_t$ is monotonic increasing function of `r' and $0<\omega_r,\omega_t<1$ i.e the underlying fluid is non-exotic in nature.\\

~~Further for a fluid sphere the trace of the energy tensor should be positive suggested by Bondi\citep{bondi99}. To check this condition for our model we have plotted $\rho-p_r-2p_t$ vs `r' in fig.~6. From the figure it is clear that $\rho-p_r-2p_t \geq 0$.

\section{Exterior Spacetime And Matching Condition}
The exterior spacetime corresponding to our interior solution is described by the schwarzschild spacetime given by the line element
\[ds^{2}=-\left(1-\frac{2M}{r}\right)dt^{2}+\left(1-\frac{2M}{r}\right)^{-1}dr^{2}\]
\begin{equation}
+r^{2}(d\theta^{2}+\sin^{2}\theta d\phi^{2})
\end{equation}

Now using the continuity of the metric coefficient $g_{tt},g_{rr},\frac{\partial g_{tt}}{\partial r}$ at the boundary $r=a$ we have,
\begin{equation}
1-\frac{2M}{a}=e^{Ba^{2}+C}
\end{equation}
\begin{equation}
\left(1-\frac{2M}{a}\right)^{-1}=e^{Aa^{2}}
\end{equation}
\begin{equation}
\frac{M}{a^{2}}=Bae^{Ba^{2}+C}
\end{equation}
Solving equations $(16)-(18)$ we get,
\begin{equation}
A=-\frac{1}{a^{2}}\ln\left(1-\frac{2M}{a}\right)
\end{equation}
\begin{equation}
B=\frac{M}{a^{3}}\left(1-\frac{2M}{a}\right)^{-1}
\end{equation}
\begin{equation}
C=\ln\left(1-\frac{2M}{a}\right)-\frac{M}{a}\left(1-\frac{2M}{a}\right)^{-1}
\end{equation}
The numerical values of the constants `A' and `B' for the strange stars SAXJ1808.4-3658(SS1)(radius=7.07 km),4U1820-30 (radius=10 km),Vela X-12 (radius=9.99 km),PSR~J 1614-2230 (radius=10.3 km) are shown in Table~I.

\section{Energy Condition}
For our anisotropic fluid sphere all the energy conditions namely Null Energy Condition (NEC),Weak Energy Condition (WEC),Strong Energy Condition (SEC) and Dominant Energy Conditions (DEC) will be satisfied if and only if the following inequalities hold simultaneously at every point inside the fluid sphere.
\begin{equation}
(i)NEC:\rho+p_r \geq 0
\end{equation}
\begin{equation}
(ii)WEC:\rho+p_r \geq 0,\rho\geq 0
\end{equation}
\begin{equation}
(iii)SEC:\rho+p_r+2p_t \geq 0,\rho+p_r\geq 0
\end{equation}
\begin{equation}
(iv)DEC:\rho>\left|p_r\right|,\rho>\left|p_t\right|
\end{equation}

\begin{figure}[htbp]
    \centering
        \includegraphics[scale=.35]{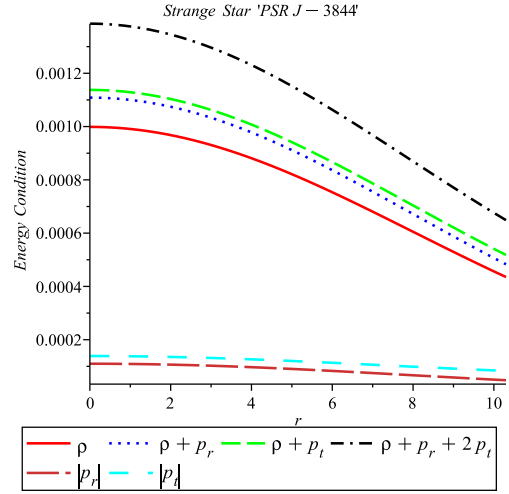}
       \caption{Energy conditions are plotted against `r' by taking $\alpha=0.11$ and $\omega_q=-0.7$ }
    \label{fig:3}
\end{figure}

We will prove the above inequalities with the help of graphical analysis.The profile of the L.H.S of the inequations (22)-(25) are shown in $fig.~7$.The figure indicates that all the above inequalities are satisfied and consequently all the energy conditions hold.

\begin{table*}
\small
\caption{The values of 'A' and 'B' obtained from the equation (20) and (21)\label{tbl-2}}
\begin{tabular}{@{}crrrrrrrrrrr@{}}
\hline
Compact Star & &$ M(M_\odot)$ && a(km) && A($km^{-2}$) && B($km^{-2}$)  \\
\hline
\hline
SAX J 1808.4-3658(SS1)& &1.435 &&7.07&&0.01823156974 &&0.014880115690 \\
4U1820-30 &&2.25 &&10&&0.01090644119 &&0.009880952381  \\
Vela~X-12&&1.77&&9.99&&0.00741034129&&0.005485958565     \\
PSR~J1614-2230\tablenotemark{a}&&1.97&&10.3&&0.00782944033&&0.006102145623\\
\hline
\end{tabular}
\tablenotetext{a}{All figures have been plotted for the strange star PSR~J1614-2230  }
\end{table*}
\section{Stability}
For a physically acceptable model of anisotropic fluid sphere one must have the radial and transverse velocity of sound should be less than 1.
Where the radial velocity $(v_{sr}^{2})$ and transverse velocity $(v_{st}^{2})$ of sound can be obtained as
\begin{equation}
v_{sr}^{2}=\frac{dp_r}{d\rho}=\alpha
\end{equation}
\begin{equation}
v_{st}^{2}=\frac{dp_t}{d\rho}
\end{equation}

\begin{figure}
    \centering
        \includegraphics[scale=.3]{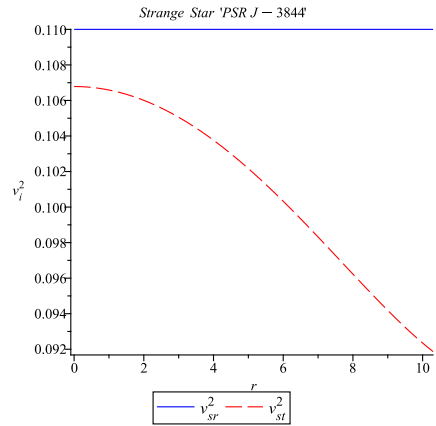}
    \caption{Radial and transverse sound velocity is plotted against `r' by taking $\alpha=0.11$ and $\omega_q=-0.7$  }
    \label{Fig 5}
\end{figure}

\begin{figure}
    \centering
        \includegraphics[scale=.3]{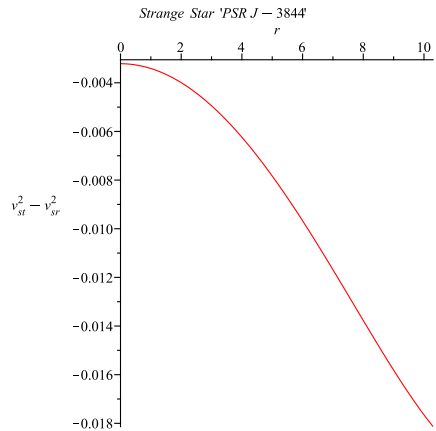}
    \caption{$v_{st}^{2}-v_{sr}^{2}$ is plotted against `r' by taking $\alpha=0.11$ and $\omega_q=-0.7$ }
    \label{Fig 5}
\end{figure}

The graphs of $(v_{sr}^{2})$ and $(v_{st}^{2})$ have been plotted for the strange star PSR~J 1614-2230 (radius=10.3 km) in Fig.~8.
From the figure it is clear that $0 <v_{sr}^{2}\leq 1$ and $0<v_{st}^{2} \leq 1$ everywhere within the stellar configuration.According to Herrera's \citep{herrera92} Cracking (or overturning) theorem for a potentially stable region one must have $v_{st}^{2}-v_{sr}^{2}<0$.From $fig.~9$ it is clear that our model satisfies this condition.So we conclude that our model is potentially stable.Moreover $0<v_{sr}^{2}\leq 1$ and $0<v_{st}^{2}<1$ therefore according to \citep{an92} $ \left|v_{st}^{2}-v_{sr}^{2}\right|\leq 1 $ which is also clear from $fig.~10$

\begin{table*}
\small
\caption{Calculated values of the parameter from the model by taking $\alpha=0.11$ \label{tbl-2}}
\begin{tabular}{@{}crrrrrrrrrrr@{}}
\hline
Compact Star  & Mass   &&Mass from &$\rho_0$ &$\rho_a$& $p_0$ \\
              &Standard Data&&the model  &(gm/cc)&(gm/cc) &$(dyne/cm^{2})$  \\
              & (in km)       &&(in km)                                        \\
\hline
\hline                                                                            \\
SAX J 1808.4-3658 &2.116625 &&~~2.093970981&~~3.203060339$\times10^{15}$&~$1.287630256\times10^{15}$ &~~3.171029736$\times10^{35}$  \\
(SS1)&   &&  &  & &\\
4U1820-30              &3.31875 &&~~3.383132691&~~2.010869426$\times10^{15}$ &~$  6.756521271\times10^{14}$&~~1.990760732$\times10^{35}$ \\
Vela~X-12 &2.61075&&~~2.521505015   &~~1.247524131$\times 10^{15}$& ~$5.950690105\times10^{14}$&~~1.235048891 $\times10^{35}$                                                                                 \\
PSR~ J1614-2230 &2.90575&&~~2.949975969&~~1.347672577 $\times 10^{15}$ &~$5.875852436\times10^{14}$&~~$ 1.334195852\times 10^{35}$ \\
\hline
\end{tabular}
\end{table*}
\begin{table*}
\small
\caption{Calculated values of the parameter from the model\label{tbl-2}by taking $\alpha=0.11$ }
\begin{tabular}{@{}crrrrrrrrrrr@{}}
\hline
Compact Star  & $\frac{M}{a}$   && $\frac{M}{a}$ from  && $ \frac{2M}{a}$ &~~~~ $(\rho_a/\rho_0)$& $z_s$  \\
              & Standard Data   &&  the model          &                 &&             & (Max Value)  \\
\hline
\hline                                                                            \\
SAX J 1808.4-3658 &0.2993811881&&~0.2961769421&&~0.5923538842$<\frac{8}{9}$ &0.402&~~0.566240125 \\
(SS1)&   &  &  & &\\
4U1820-30&0.3318750000&&~0.3383132691&&~0.6766265382$<\frac{8}{9}$&0.336              &~~0.758522025 \\
Vela~X-12 &0.2613363363&&~0.2524029044&&~0.5048058088$<\frac{8}{9}$&0.477 &~~0.421059391                                                                                \\
PSR~ J1614-2230  &0.2821116505&&~0.2864054339&&~0.2864054339$<\frac{8}{9}$&0.436&~~0.529994615\\
\hline
\end{tabular}
\end{table*}

\begin{figure}
    \centering
        \includegraphics[scale=.3]{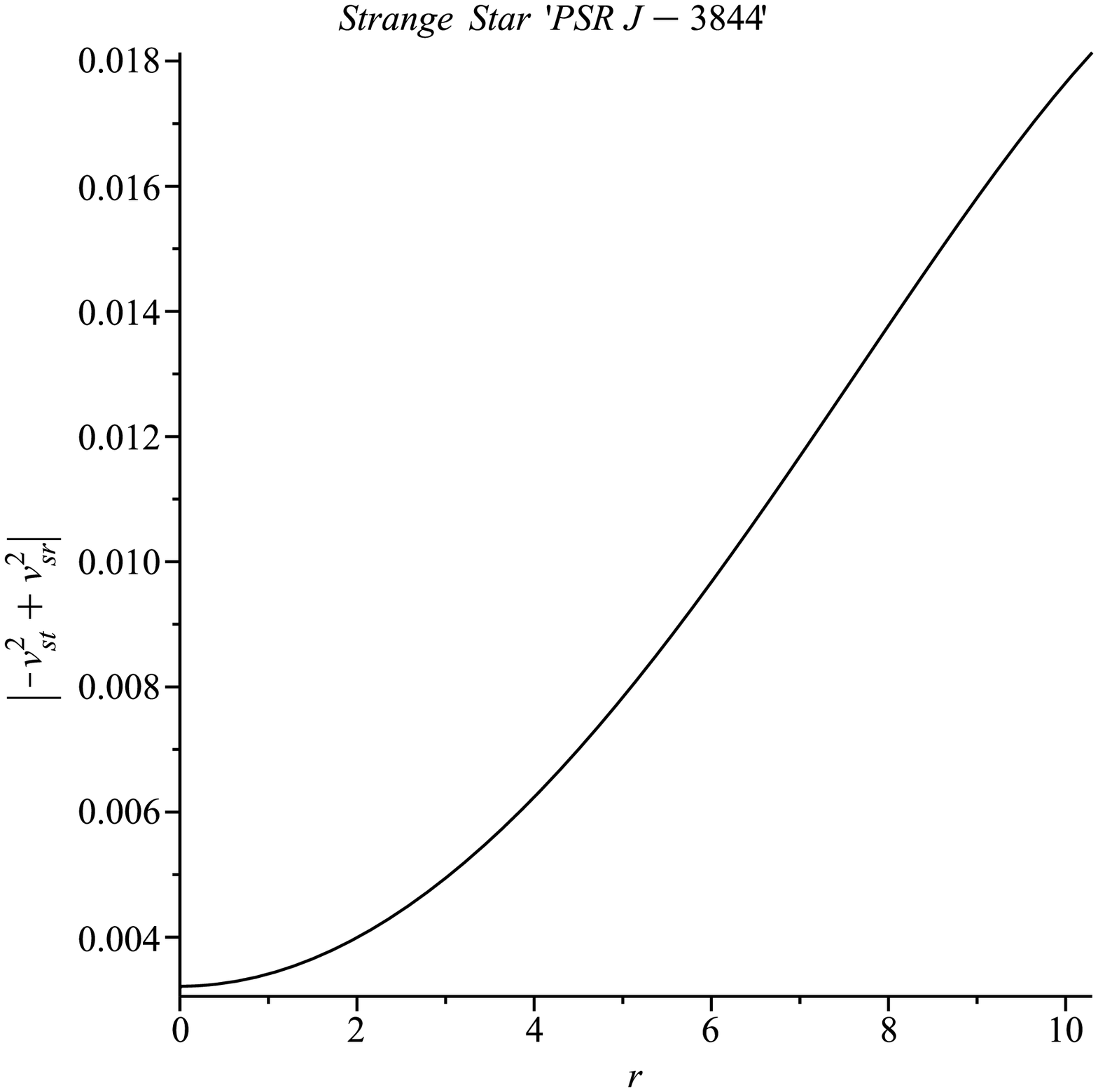}
    \caption{$\left|v_{st}^{2}-v_{sr}^{2}\right|$ is plotted against `r' by taking $\alpha=0.11$ and $\omega_q=-0.7$ }
    \label{Fig 5}
\end{figure}

\section{Some Features}

\subsection{Mass Function}
The mass function within the radius `r'can be obtained as,

\[m(r)=\int_0^{r}4\pi \rho r^{2}dr\]
\begin{equation}
=\frac{A+B}{2A(1+m)}\left[\frac{1}{2}\sqrt{\frac{\pi}{A}}erf(\sqrt{A}r)-re^{-Ar^{2}}\right]
\end{equation}

As $r\rightarrow 0$ ,$m(r)\rightarrow 0$,i.e, the mass function is regular at the origin.The profile of mass function for strange star PSR~J 1614-2230 (10.3 km) against `r' is shown in fig.~11. The figure shows that $m(r)>0$ for $0<r<a$ and is monotonic increasing function of `r'. We have calculated the values of mass for the strange stars SAXJ1808.4-3658(SS1)(radius=7.07 km),4U1820-30 (radius=10 km),Vela X-12 (radius=9.99 km),PSR~J 1614-2230 (radius=10.3 km) from our model which is given in Table II and have also compared the values of mass of those strange stars to the observational data (see Table II)

\begin{figure}
    \centering
        \includegraphics[scale=.3]{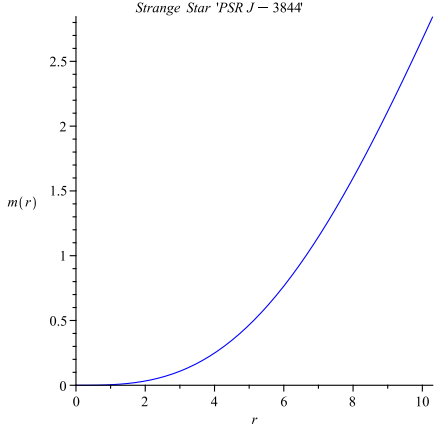}
    \caption{mass function is plotted against `r' by taking $\alpha=0.11$ and $\omega_q=-0.7$ }
    \label{Fig 5}
\end{figure}

\subsection{Compactness}
The compactness of the star $u(r)$ can be defined by,
\begin{equation}
u(r)=\frac{m(r)}{r}=\frac{A+B}{2A(1+m)}\left[\frac{1}{2}\sqrt{\frac{\pi}{A}}\frac{erf(\sqrt{A}r)}{r}-e^{-Ar^{2}}\right]
\end{equation}
The figure of u(r) for the strange star PSR~J 1614-2230 (radius=10.3 km) is depicted in $fig.~12$
\begin{figure}
    \centering
        \includegraphics[scale=.3]{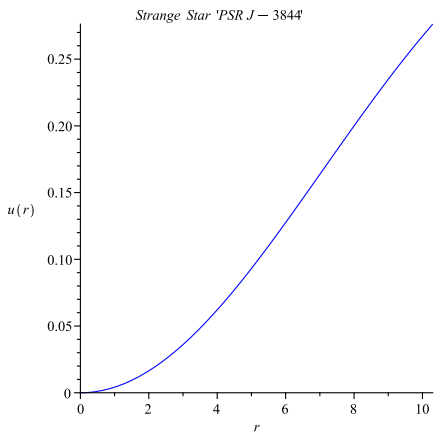}
    \caption{The compactness of the strange star is plotted against `r' by taking $\alpha=0.11$ and $\omega_q=-0.7$  }
    \label{Fig 5}
\end{figure}

\subsection{Mass Radius Relation}
In this section we will discuss about the mass radius relation of the strange stars SAXJ1808.4-3658(SS1)(radius =7.07 km),4U1820-30 (radius=10 km),Vela X-12 (radius=9.99 km),PSR~J 1614-2230 (radius=10.3 km). According to Buchdahl \citep{buchdahl59} twice the maximum allowable ratio of Mass to the radius for an astrophysical object lies in the range $\frac{2m}{a}<\frac{8}{9}$ where `m' is the mass and `a' is the radius of the object. We have calculated the values of $\frac{2m}{a}$ for the strange stars SAXJ1808.4-3658(SS1)(radius=7.07 km),4U1820-30 (radius=10 km),Vela X-12 (radius=9.99 km),PSR~J 1614-2230 (radius=10.3 km) from our model which is given in Table III.From the table it is clear that calculated values of $\frac{2m}{a}$ of the strange stars SAXJ1808.4-3658(SS1)(7.07 km),4U1820-30 (10 km),Vela X-12 (9.99 km),PSR~J 1614-2230 (10.3 km) lie in the expected range. In this connection we want to mention that $\frac{m}{a}$ is called the `compactification factor' which classifies the stellar object as follows \citep{jotania06}:

\begin{itemize}
  \item for normal star $\frac{m}{a}\sim 10^{-5}$
  \item for white dwarfs $\frac{m}{a}\sim 10^{-3}$
  \item for neutron star $10^{-1}<\frac{m}{a}<\frac{1}{4}$
  \item for ultracompact star $\frac{1}{4}<\frac{m}{a}<\frac{1}{2}$
  \item for blackhole $\frac{m}{a}=\frac{1}{2}$
\end{itemize}

and the compactification parameter for strange stars should lie in the range of ultracompact stars that have the matter density beyond the nuclear density ($\sim 10^{14}$ gm/cc). The calculated values of the compactification factor of the strange stars SAXJ1808.4-3658(SS1)(radius=7.07 km),4U1820-30 (radius=10 km), Vela X-12 (radius=9.99 km),PSR~J 1614-2230 (radius=10.3 km) from our model is shown in Table III. From the table it is clear that the values of $\frac{m}{a}$ of these strange stars lies between $\frac{1}{4}$ and $\frac{1}{2}$.

\subsection{Surface Redshift}
The redshift function $z_s$ can be obtained as,
\[z_s=(1-2u)^{-\frac{1}{2}}-1\]
\begin{equation}
=\left[1-\frac{A+B}{A(1+m)}\left(\frac{1}{2}\sqrt{\frac{\pi}{A}}\frac{erf(\sqrt{A}r)}{r}-e^{-Ar^{2}}\right)\right]^{-\frac{1}{2}}-1
\end{equation}
The profile of the redshift function of the strange star PSR~J 1614-2230 (radius=10.3 km) is shown in fig.~13.In this connection we want to mentioned that for anisotropic star the values of the maximum surface redshift cannot be arbitrarily large. On the other hand,according to  B\"{o} hmer and Harko for an anisotropic star in the presence of a cosmological constant the surface redshift should lie in the range $z_s\leq 5 $ \citep{harko06}. The maximum values of the surface redshift of the strange stars SAXJ1808.4-3658(SS1)(radius=7.07 km),4U1820-30 (radius=10 km),Vela X-12 (radius=9.99 km),PSR~J 1614-2230 (radius=10.3 km) calculated from our model are shown in Table III. Though our model is without cosmological constant from the table it is clear that maximum value of $z_s$ of the above mentioned strange stars is less than 5, which is quite reasonable.

\section{Junction Condition}
To avoid the singular behavior of the physical variable our solution should satisfy the Darmois conditions on the boundary \citep{herrera08}. Such conditions require the continuity of the first and second fundamental form across the boundary. According to Bonnor \& Vickers \citep{bonnor81} these conditions are equivalent to the conditions imposed by Lichnerowicz (Theories Relativistes de la Gravitation et de l'Electromagnetisme (Masson, Paris, 1955)).These conditions require (for the smooth matching) the existence of a coordinate system where the metric and all their first derivatives are continuous across the boundary surface. If the second fundamental form is not continuous, then there is a shell on the boundary surface, and the matching is described by the Israel conditions. Now it is obvious from equation (12) that the radial pressure is non-vanishing at the boundary. So according to Misner and Sharp, \citep{misner64} the second fundamental form is discontinuous across the boundary surface. To take care of this let us use the Darmois-Israel \citep{isreal66,isreal67} formation to determine the surface stresses at the junction boundary. The intrinsic surface stress energy tensor $S_{ij}$ is given by Lancozs equations in the following form    \\

\begin{equation}
S^{i}_{j}=-\frac{1}{8\pi}(K^{i}_j-\delta^{i}_j K^{k}_k)
\end{equation}

\begin{figure}
    \centering
        \includegraphics[scale=.3]{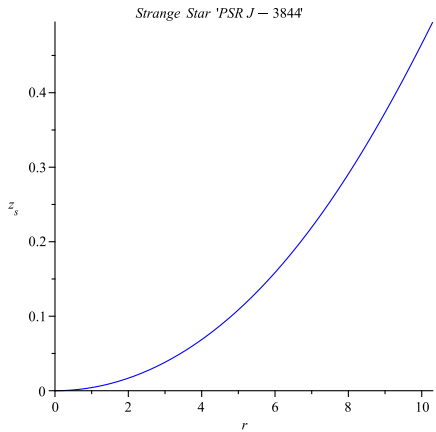}
    \caption{The variation of the surface redshift is shown against `r' by taking $\alpha=0.11$ and $\omega_q=-0.7$ }
    \label{Fig 5}
\end{figure}

The discontinuity in the second fundamental form is given by,
\begin{equation}
K_{ij}=K_{ij}^{+}-K_{ij}^{-}
\end{equation}
where the second fundamental form is given by,
\begin{equation}
K_{ij}^{\pm}=-n_{\nu}^{\pm}\left[\frac{\partial^{2}X_{\nu}}{\partial \xi^{1}\partial\xi^{j}}+\Gamma_{\alpha\beta}^{\nu}\frac{\partial X^{\alpha}}{\partial \xi^{i}}\frac{\partial X^{\beta}}{\partial \xi^{j}} \right]|_S
\end{equation}

where $n_{\nu}^{\pm}$ are the unit normal vector defined by,

\begin{equation}
n_{\nu}^{\pm}=\pm\left|g^{\alpha\beta}\frac{\partial f}{\partial X^{\alpha}}\frac{\partial f}{\partial X^{\beta}}  \right|^{-\frac{1}{2}}\frac{\partial f}{\partial X^{\nu}}
\end{equation}
with $n^{\nu}n_{\nu}=1$.Where $\xi^{i}$ is the intrinsic coordinate on the shell.$+$ and $-$ corresponds to exterior i.e, Schwarzschild  spacetime and interior (our) spacetime respectively.\\
Considering the spherical symmetry of the spacetime surface stress energy tensor can be written as $S^{i}_j=diag(-\sigma,\mathcal{P})$.Where $\sigma$ and $\mathcal{P}$ is the surface energy density and surface pressure respectively.

The non-trivial components of the extrinsic curvature are given by
\begin{equation}
K_{\tau}^{\tau+}=\frac{\frac{M}{a^{2}}+\ddot{a}}{\sqrt{1-\frac{2M}{a}+\dot{a}^{2}}}
\end{equation}
\begin{equation}
K_{\tau}^{\tau-}=\frac{-aAe^{-Aa^{2}}+\ddot{a}}{\sqrt{e^{-Aa^{2}}+\dot{a}^{2}}}
\end{equation}
and
\begin{equation}
K_{\theta}^{\theta+}=\frac{1}{a}\sqrt{1-\frac{2M}{a}+\dot{a}^{2}}
\end{equation}
\begin{equation}
K_{\theta}^{\theta-}=\frac{1}{a}\sqrt{e^{-Aa^{2}}+\dot{a}^{2}}
\end{equation}
Therefore the surface stress energy and surface pressure is given by,

\begin{equation}
\sigma=-\frac{1}{4\pi a}\left[\sqrt{1-\frac{2M}{a}}-e^{-\frac{Aa^{2}}{2}}\right]
\end{equation}
\begin{equation}
\mathcal{P}=\frac{1}{8\pi a}\left[\frac{1-\frac{M}{a}}{\sqrt{1-\frac{2M}{a}}}-(1+2Ba^{2})e^{-\frac{Aa^{2}}{2}}\right]
\end{equation}
Where $\sigma$ and $\mathcal{P}$ are respectively the surface stress energy and surface pressure.\\
Hence one can match our interior spacetime to the exterior Schwarzschild spacetime in presence of a thin shell.

The mass of the thin shell is given by
\begin{equation}
m_s=4\pi a^{2}\sigma
\end{equation}
Using (39) and (41) one can obtain the mass of the fluid sphere as,

\begin{equation}
M=\frac{a}{2}\left(1-e^{-Aa^{2}}\right)-\frac{m_s^{2}}{2}+m_se^{-\frac{Aa^{2}}{2}}
\end{equation}
Which gives the mass of the fluid sphere in terms of the thin shell mass.\\

~~~~Next we will discuss about the evolution identity given by \[[T_{\mu\nu }n^{\mu}n^{\nu}]_{-}^{+}=\overline{K}^{i}_{j}S^{j}_{i}\]
where
\begin{equation}
\overline{K}^{i}_{j}=\frac{1}{2}\left(K_{j}^{i+}+K_{j}^{i-}\right)
\end{equation}
 which gives,

\[ p_r+\frac{(\rho+p_r)\dot{a}^{2}}{e^{-Aa^{2}}}\]
\[ =-\frac{1}{2a}\left(\sqrt{1-\frac{2M}{a}+\dot{a}^{2}}+\sqrt{e^{-Aa^{2}}+\dot{a}^{2}}\right)\mathcal{P}\]
 \begin{equation}
~~~~ +\frac{1}{2}\left(\frac{\frac{M}{a^{2}}+\ddot{a}}{\sqrt{1-\frac{2M}{a}+\dot{a}^{2}}}
 +\frac{-aAe^{-Aa^{2}}+\ddot{a}}{\sqrt{e^{-Aa^{2}}+\dot{a}^{2}}}\right)\sigma
\end{equation}
For a static solution $a_0$ (assuming $\dot{a}=0=\ddot{a}$) from the above equation one can obtain

\[p_r(a_0)=-\frac{1}{2a_0}\left(\sqrt{1-\frac{2M}{a_0}}+e^{-\frac{Aa_0^{2}}{2}}\right)\mathcal{P}\]
\begin{equation}
~~~~~+\frac{1}{2}\left(\frac{\frac{M}{a_0^{2}}}{\sqrt{1-\frac{2M}{a_0}}}-a_0Ae^{-\frac{Aa_0^{2}}{2}}\right)\sigma
\end{equation}
Now in particular if the surface energy density $(\sigma)$ vanishes from equation (39) and (45) one obtain
\begin{equation}
p_r(a_0)=-\frac{1}{a_0}\left(\sqrt{1-\frac{2M}{a_0}}\right)\mathcal{P}
\end{equation}
Since radial pressure $(p_r)$ is positive (see fig.~2) for our model then from equation (46) it is clear that surface pressure $\mathcal{P}<0$ is required to holds the thin shell again collapse.

\section{Discussion and concluding remarks}
In the present paper we have proposed a new model of anisotropic star using KB ansatz \citep{kb75} in presence of quintessence field which is characterized by a parameter $\omega_q$ with $1-<\omega_q<-\frac{1}{3}.$ Inspired by the observational evidence of the accelerated expansion of our present universe we have considered quintessence dark energy field in our present stellar model.Our present model is free from central singularity.Both the radial pressure $(p_r)$ and matter density $(\rho)$ are monotonic decreasing function of `r',i.e,they have maximum value at the center and it decreases from the center to the surface of the star. Plugging `G' and `c' in the expression of $\rho,p_r$ given in equation (11) and (12)we have obtained the values of central and surface density and central pressure from our model which are given in Table 2.From the table it is clear that central density $\rho_0$ of different strange stars are $\sim 10^{15} gm/cc$ i.e beyond the nuclear density.  All the energy conditions are satisfied by our model. The subliminal velocity of sound is less than 1,i.e, $0<v_{sr}^{2},v_{st}^{2}<1$ and $v_{sr}^{2}>v_{st}^{2}$.So our model is potentially stable. The usual ratio of twice mass to the radius calculated for the different strange star SAXJ1808.4-3658(SS1)(radius=7.07 km),4U1820-30 (radius=10 km),Vela X-12 (radius=9.99 km),PSR~J 1614-2230 (radius=10.3 km)  $<\frac{8}{9}$, lies in the range proposed by Buchdahl \citep{buchdahl59}. We have calculated the numerical values of the mass of above mentioned strange star from our model which is very close to the observational data. The maximum values of the surface redshift function is calculated for the strange stars from our model which is less than 5 \citep{harko06}. Our interior solutions is matched to the exterior schwarzschild line element in presence of thin shell and we have shown that negative surface pressure is required to hold the system against collapse. A relation among radial pressure $(p_r)$, surface energy density $(\sigma)$,surface pressure ($\mathcal{P}$) is also obtained.

\section{Acknowledgement }

PB is grateful to the anonymous referee for his/her valuable comments to improve the manuscript.

\end{document}